\documentclass[12pt]{article}
\usepackage[dvips]{color}
\usepackage{epsfig}
\usepackage{amsmath}
\usepackage{graphicx}
\usepackage{amssymb}
\usepackage{float}

\begin{document}
\title{\bf Analytical Calculation of Stored Electrostatic Energy per Unit
Length for an Infinite Charged Line and an Infinitely Long
Cylinder in the Framework of Born-Infeld Electrostatics}

\author{S. K. Moayedi \thanks{E-mail:
s-moayedi@araku.ac.ir}\hspace{1mm} , M. Shafabakhsh
\thanks{E-mail: m-shafabakhsh@phd.araku.ac.ir}\hspace{1mm},
 F. Fathi \thanks{E-mail: f-fathi@phd.araku.ac.ir}\hspace{1.5mm}  \\
{\small {\em  Department of Physics, Faculty of Sciences,
Arak University, Arak 38156-8-8349, Iran}}\\
}
\date{\small{}}
\maketitle

\begin{abstract}
\noindent More than 80 years ago, Born-Infeld electrodynamics was
proposed in order to remove the point charge singularity in
Maxwell electrodynamics. In this work, after a brief introduction
to Lagrangian formulation of Abelian Born-Infeld model in the
presence of an external source, we obtain the explicit forms of
Gauss's law and the energy density of an electrostatic field for
Born-Infeld electrostatics. The electric field and the
stored electrostatic energy per unit length for an infinite
charged line and an infinitely long cylinder in Born-Infeld
electrostatics are calculated. Numerical estimations in this
paper show that the nonlinear corrections to Maxwell
electrodynamics are considerable only for strong electric fields.
We present an action functional for Abelian Born-Infeld model
with an auxiliary scalar field in the presence of an external
source. This action functional is a generalization of the action
functional which was presented by Tseytlin in his studies on low
energy dynamics of $D$-branes (Nucl. Phys. \textbf{B469}, 51
(1996); Int. J. Mod. Phys. A \textbf{19}, 3427 (2004)). Finally,
we derive the symmetric energy-momentum tensor for Abelian
Born-Infeld model with an auxiliary scalar field.

\noindent
\hspace{0.35cm}



\end{abstract}

\section{Introduction}
Maxwell electrodynamics is a very successful theory which
describes a wide range of macroscopic phenomena in electricity and
magnetism. On the other hand, in Maxwell electrodynamics, the
electric field of a point charge $q$ at the position of the point
charge is singular, i.e., $$ \textbf E(\textbf
x)=\frac{q}{4\pi\epsilon_{0}|\textbf x|^{2}}\frac{\textbf
x}{|\textbf x|}\; \stackrel{|\textbf
x|\rightarrow0}{-\!\!-\!\!\!\longrightarrow} \infty. $$ Also, in
Maxwell electrodynamics, the classical self-energy of a point
charge is $$
U=\frac{q^2}{8\pi\epsilon_{0}}\int^{\infty}_{0}\frac{dr}{r^2}\rightarrow\infty.$$
More than 80 years ago, Born and Infeld proposed a nonlinear
generalization of Maxwell electrodynamics [1]. In their
generalization, the classical self-energy of a point charge was a
finite value [1-6]. Recent studies in string theory show that the
dynamics of electromagnetic fields on $D$-branes can be described
by Born-Infeld theory [7-10]. In a paper on Born-Infeld theory
[8], the concept of a BIon was introduced by Gibbons. BIon is a
finite energy solution of a nonlinear theory with a
distributional source. Today, many physicists believe that the
dark energy in our universe can be described by a Born-Infeld type
scalar field [11]. The authors of Ref. [12] have presented a
non-Abelian generalization of Born-Infeld theory. In their
generalization, they have found a one-parameter family of finite
energy solutions in the case of the $SU(2)$ gauge group. In 2013,
Hendi [13] proposed a nonlinear generalization of Maxwell
electrodynamics which is called exponential electrodynamics
[14,15]. The black hole solutions of Einstein's gravity in the
presence of exponential electrodynamics in a $3+1$-dimensional
spacetime are obtained in Ref. [13]. In 2014, Gaete and
Helayel-Neto introduced a new generalization of Maxwell
electrodynamics which is known as logarithmic electrodynamics
[14]. They proved that the classical self-energy of a point
charge in logarithmic electrodynamics is a finite value.
Recently, a novel generalization of Born-Infeld electrodynamics
is presented by Gaete and Helayel-Neto in which the authors show
that the field energy of a point-like charge is finite only for
Born-Infeld like electrodynamics [15]. In Ref. [16], a nonlinear
model for electrodynamics with manifestly broken gauge symmetry
is proposed. In the above mentioned model for nonlinear
electrodynamics, there are non-singular solitonic solutions which
describe charged particles. Another interesting theory of
nonlinear electrodynamics was proposed and developed by Heisenberg and his students Euler and Kockel [17-19]. They showed
that classical electrodynamics must be corrected by nonlinear
terms due to the vacuum polarization effects. In Ref. [20], the
charged black hole solutions for Einstein-Euler-Heisenberg theory
are obtained. There are three physical motivations in writing
this paper. First, the exact solutions of nonlinear field
equations are very important in theoretical physics. These
solutions help us to obtain a better understanding of physical
reality. According to above statements, we attempt to obtain
particular cylindrically symmetric solutions in Born-Infeld
electrostatics. The search for spherically symmetric solutions in
Born-Infeld electrostatics will be discussed in future works.
Second, we want to show that the nonlinear corrections in
electrodynamics are considerable only for very strong electric
fields and extremely short spatial distances. Third, we hope to
remove or at least modify the infinities which appear in
Maxwell electrostatics. This paper is organized as follows. In
Section 2, we study Lagrangian formulation of Abelian Born-Infeld
model in the presence of an external source. The explicit forms
of Gauss's law and the energy density of an electrostatic field
for Born-Infeld electrostatics are obtained. In Section 3, we
calculate the electric field together with the stored
electrostatic energy per unit length for an infinite charged line
and an infinitely long cylinder in Born-Infeld electrostatics.
Summary and conclusions are presented in Section 4. Numerical
estimations in Section 4 show that the nonlinear corrections to
electric field of an infinite charged line at large radial
distances are negligible for weak electric fields. There are two
appendices in this paper. In Appendix A, a generalized action
functional for Abelian Born-Infeld model with an auxiliary scalar
field in the presence of an external source is proposed. In
Appendix B, we obtain the symmetric energy-momentum tensor for
Abelian Born-Infeld model with an auxiliary scalar field. We use
SI units throughout this paper. The metric of spacetime has the
signature $(+,-,-,-)$.

\section{Lagrangian Formulation of Abelian Born-Infeld Model with an External Source}

The Lagrangian density for Abelian Born-Infeld model in a
$3+1$-dimensional spacetime is [1-6]
\begin{equation}
{\cal L}_{_{BI}}=\epsilon_{0}\beta^{2}\bigg\lbrace
1-\sqrt{1+\frac{c^{2}}{2\beta^{2}}F_{\mu\nu}F^{\mu\nu}}\bigg\rbrace
-J^{\mu}A_{\mu},
\end{equation}
where $F_{\mu\nu}=\partial_{\mu}A_{\nu}-\partial_{\nu}A_{\mu}$ is
the electromagnetic field tensor and $J^{\mu}=(c \rho,\textbf J)$
is an external source for the Abelian field $A^{\mu}=(\frac{1}{c}
\phi , \textbf A)$. The parameter $\beta$ in Eq. (1) is called
the nonlinearity parameter of the model. In the limit
$\beta\rightarrow\infty$, Eq. (1) reduces to the Lagrangian
density of the Maxwell field, i.e.,
\begin{equation}
{\cal L}_{_{BI}}|_{_{large \; \beta}}={\cal L}_{_{M}}+{\cal
O}(\beta^{-2}),
\end{equation}
where ${\cal
L}_{_{M}}=-\frac{1}{4\mu_{0}}F_{\mu\nu}F^{\mu\nu}-J^{\mu}A_{\mu}$
is the Maxwell Lagrangian density. The Euler-Lagrange equation
for the Born-Infeld field $A^{\mu}$ is
\begin{equation}
\frac{\partial{\cal L}_{_{BI}}}{\partial
A_{\lambda}}-\partial_{\rho}\bigg(\frac{\partial{\cal
L}_{_{BI}}}{\partial(\partial_{\rho}A_{\lambda})}\bigg)=0.
\end{equation}
If we substitute Lagrangian density (1) in the Euler-Lagrange
equation (3), we will obtain the inhomogeneous Born-Infeld
equations as follows:
\begin{equation}
\partial_{\rho}\bigg(\frac{F^{\rho\lambda}}{\sqrt{1+\frac{c^{2}}{2\beta^{2}}F_{\mu\nu}F^{\mu\nu}}}\bigg)=\mu_{0}J^{\lambda}.
\end{equation}
The electromagnetic field tensor $F_{\mu\nu}$ satisfies the
Bianchi identity:
\begin{equation}
\partial_{\mu}F_{\nu\lambda}+\partial_{\nu}F_{\lambda\mu}+\partial_{\lambda}F_{\mu\nu}=0.
\end{equation}
Equation (5) leads to the homogeneous Maxwell equations. In
$3+1$-dimensional spacetime, the components of $F_{\mu\nu}$ can be
written as follows:
\begin{eqnarray}
F_{\mu\nu}&=& \left( {\begin{array}{cccc}
   0 & {E_{x}}/{c}\ & {E_{y}}/{c}  & {E_{z}}/{c}  \\
   -{E_{x}}/{c} & 0 & -B_{z} &B _{y} \\
   -{E_{y}}/{c} & B_{z} & 0 & -B_{x} \\
   -{E_{z}}/{c} & -B_{y} & B_{x} & 0
    \end{array}} \right).
\end{eqnarray}
Using Eq. (6), Eqs. (4) and (5) can be written in the vector form
as follows:
\begin{eqnarray}
\boldsymbol{\nabla} \cdot\bigg(\frac{\textbf E(\textbf x
,t)}{\sqrt{1-\frac{\textbf E^{2}(\textbf x ,t)-c^{2}\textbf
B^{2}(\textbf x ,t)}{\beta^{2}}}}\bigg) &=& \frac{\rho(\textbf x
,t)}{\epsilon_{0}}, \\
 \boldsymbol{\nabla} \times \textbf E(\textbf x ,t) &=& -\frac{\partial
\textbf B(\textbf x ,t)}{\partial t}, \\
\boldsymbol{\nabla} \times \bigg(\frac{\textbf B(\textbf x
,t)}{\sqrt{1-\frac{\textbf E^{2}(\textbf x ,t)-c^{2}\textbf
B^{2}(\textbf x ,t)}{\beta^{2}}}}\bigg) &=& \mu_{0}\ \textbf J
(\textbf x ,t)+\frac{1}{c^{2}} \frac{\partial}{\partial
t}\bigg(\frac{\textbf E(\textbf x ,t)}{\sqrt{1-\frac{\textbf
E^{2}(\textbf x
,t)-c^{2}\textbf B^{2}(\textbf x ,t)}{\beta^{2}}}}\bigg), \\
\boldsymbol{\nabla} \cdot \textbf B (\textbf x ,t) &=& 0 .
\end{eqnarray}
The symmetric energy-momentum tensor for the Abelian Born-Infeld
model in Eq. (1) has been obtained by Accioly [21] as follows:
\begin{equation}
T^{\mu}_{\
\lambda}=\frac{1}{\mu_{0}}\bigg[\frac{F^{\mu\nu}F_{\nu\lambda}}{\Omega}+\frac{\beta^{2}}{c^{2}}(\Omega-1)\delta^{\mu}_{\
\lambda}\bigg],
\end{equation}
where
$\Omega:=\sqrt{1+\frac{c^{2}}{2\beta^{2}}F_{\alpha\gamma}F^{\alpha\gamma}}$.
The classical Born-Infeld equations (7)-(10) for an electrostatic
field $\textbf E(\textbf x)$ are
\begin{eqnarray}
\boldsymbol{\nabla} \cdot\bigg(\frac{\textbf E(\textbf
x)}{\sqrt{1-\frac{\textbf E^{2}(\textbf x)}{\beta^{2}}}}\bigg) &=&
\frac{\rho(\textbf x )}{\epsilon_{0}},  \\
\boldsymbol{\nabla} \times \textbf E (\textbf x ) &=& 0 .
\end{eqnarray}
Equations (12) and (13) are fundamental equations of Born-Infeld
electrostatics [2]. Using divergence theorem, the integral form of
Eq. (12) can be written in the form
\begin{equation}
\oint_{S}\frac{1}{\sqrt{1-\frac{\textbf E^{2}(\textbf
x)}{\beta^{2}}}}\;\textbf E(\textbf x).\textbf{n}\; da
=\frac{1}{\epsilon_{0}} \int_{V}  \rho (\textbf x) d^{3}x ,
\end{equation}
where $V$ is the three-dimensional volume enclosed by a
two-dimensional surface $S$. Equation (14) is Gauss's law in
Born-Infeld electrostatics. Using Eqs. (6) and (11), the energy
density of an electrostatic field in Born-Infeld theory is given
by
\begin{equation}
u(\textbf x )=\epsilon_{0}\beta^{2}
\bigg(\frac{1}{\sqrt{1-\frac{\textbf E^{2}(\textbf
x)}{\beta^{2}}}}-1\bigg).
\end{equation}
In the limit $\beta\rightarrow\infty$, the modified electrostatic
energy density in Eq. (15) smoothly becomes the usual
electrostatic energy density in Maxwell theory, i.e.,
\begin{equation}
u(\textbf x )|_{_{large \; \beta}}=\frac{1}{2}\epsilon_{0}\textbf
E^{2}(\textbf x)+{\cal O}(\beta^{-2}).
\end{equation}

\section{Calculation of Stored Electrostatic Energy per Unit
 Length for an Infinite Charged Line and an Infinitely Long Cylinder in Born-Infeld Electrostatics}

\subsection{Infinite Charged Line}

Let us consider an infinite charged line with a uniform positive
linear charge density $\lambda$ which is located on the $z$-axis.
Now, we find an expression for the electric field $\textbf
E(\textbf x)$ at a radial distance $\rho$ from the $z$-axis.
Because of the cylindrical symmetry of the problem, the suitable
Gaussian surface is a circular cylinder of radius $\rho$ and
length $L$, coaxial with the $z$-axis (see Fig. 1).

\begin{figure}[h]
\centerline{\includegraphics[width=4 cm]{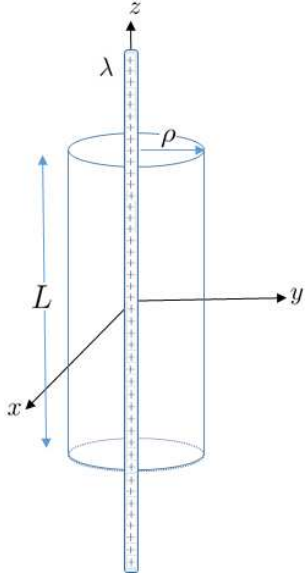}} \caption{\small
The Gaussian surface for an infinite charged line. The cylindrical
symmetry of the problem implies that $\textbf{E}(\textbf{x})=
E_{\rho}(\rho)\,\hat{\textbf{e}}_{\rho}$, where
$\hat{\textbf{e}}_{\rho}$ is the radial unit vector in cylindrical
coordinates$(\rho,\varphi,z)$.}
\end{figure}

Using the cylindrical symmetry of the problem together with the
modified Gauss's law in Eq. (14), the electric field for the
Gaussian surface in Fig. 1 becomes
\begin{equation}
\textbf{E}(\textbf{x})=
\dfrac{\lambda}{2\pi\epsilon_{0}\rho}\dfrac{1}{\sqrt{1+\big(\frac{\lambda}{2\pi\epsilon_{0}\beta
\rho}\big )^{2}}}\,\hat{\textbf{e}}_{\rho}.
\end{equation}
 In contrast with Maxwell electrostatics, the
electric field $\textbf{E}(\textbf{x})$ in Eq. (17) has a finite
value on the $z$-axis, i.e.,
\begin{equation}
\lim_{\rho\,\rightarrow
0}\textbf{E}(\textbf{x})=\beta\,\hat{\textbf{e}}_{\rho}.
\end{equation}
At large radial distances from the $z$-axis, the asymptotic behavior
of the electric field in Eq. (17) is given by
\begin{equation}
\textbf{E}(\textbf{x})=\dfrac{\lambda}{2\pi\epsilon_{0}\rho}\,\hat{\textbf{e}}_{\rho}-ý
\dfrac{\lambda^{3}}{16\pi^{3}\epsilon_{0}^{3}\beta^{2}\rho^{3}}\,\hat{\textbf{e}}_{\rho}+{\cal
O}(\rho^{-5}).
\end{equation}
The first term on the right-hand side of Eq. (19) shows the electric
field of an infinite charged line in Maxwell electrostatics, while
the second and higher order terms in Eq. (19) show the effect of
nonlinear corrections. By putting Eq. (17) in Eq. (15), the
electrostatic energy density for an infinite charged line in
Born-Infeld electrostatics can be written as follows:
\begin{equation}
u(\textbf{x})=\epsilon_{0}\beta^{2}\Big(\sqrt{1+\big(\dfrac{\lambda}{2\pi
\epsilon_{0}\beta\rho}\big)^{2}}-1\Big).
\end{equation}
Using Eq. (20), the stored electrostatic energy per unit length for
an infinite charged line in the radial interval $0\leq \rho \leq
\Lambda$ is given by
\begin{eqnarray}
\dfrac{U}{L}&=&\int^{\Lambda}_{0}\int^{2\pi}_{0}u(\textbf{x})\rho
d\rho d\varphi \nonumber \\  \qquad &=&
2\pi\epsilon_{0}\beta^{2}\bigg\lbrace
\dfrac{\Lambda}{2}\sqrt{\Lambda^{2}+\big(\dfrac{\lambda}{2\pi
\epsilon_{0}\beta}\big)^{2}}-\dfrac{\Lambda^{2}}{2} \nonumber \\
&+&\dfrac{1}{2}\big(\dfrac{\lambda}{2\pi\epsilon_{0}\beta}\big)^{2}\ln
\bigg(\dfrac{\Lambda+\sqrt{\Lambda^{2}+\big(\dfrac{\lambda}{2\pi
\epsilon_{0}\beta}\big)^{2}}}{\big(\dfrac{\lambda}{2\pi\epsilon_{0}\beta}\big)}\bigg)\bigg\rbrace.
\end{eqnarray}
It is necessary to note that the above value for $\frac{U}{L}$ has
an infinite value in Maxwell theory. In the limit of large
$\beta$, the expression for $\frac{U}{L}$ in Eq. (21) diverges
logarithmically as $\ln \beta$. Hence, it seems that the finite
regularization parameter $\beta$ removes the logarithmic
divergence in Eq. (21).

\subsection{Infinitely Long Cylinder}
In this subsection, we determine the electric field
$\textbf{E}(\textbf{x})$ and stored electrostatic energy per unit
length for an infinitely long cylinder of radius $R$ and uniform
positive volume charge density $\tau$. As in the previous
subsection, we assume that the Gaussian surface is a cylindrical
closed surface of radius $\rho$ and length $L$ with a common axis
with
the infinitely long cylinder (see Fig. 2).\\
\begin{figure}[h]
\centerline{\includegraphics[width=8 cm]{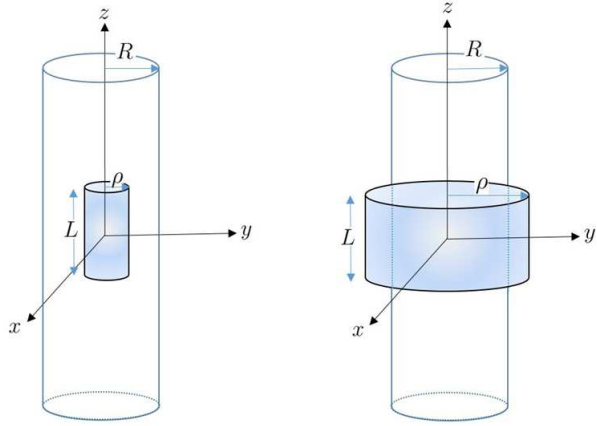}} \caption{\small
The Gaussian surface for an infinitely long cylinder of radius $R$.
Left) inside the cylinder $(\rho<R)$. Right) outside the cylinder
$(\rho>R)$.}
\end{figure}

According to modified Gauss's law in Eq. (14), the electric field
for the Gaussian surfaces in Fig. 2 is given by
\begin{equation}
\textbf{E}(\textbf{x})=
\begin{cases}
\dfrac{\tau\rho}{2\epsilon_{0}}\dfrac{1}{\sqrt{1+\big(\dfrac{\tau\rho}{2\epsilon_{0}\beta}\big)^{2}}}\,\hat{\textbf{e}}_{\rho}\;;\;\;\;\;\;\;&\rho < R,\\
\dfrac{\tau R^{2}
}{2\epsilon_{0}\rho}\dfrac{1}{\sqrt{1+\big(\dfrac{\tau
R^{2}}{2\epsilon_{0}\beta\rho}\big)^{2}}}\,\hat{\textbf{e}}_{\rho}
\;;\;\;\;\;&\rho
> R.
\end{cases}
\end{equation}

For the large values of the nonlinearity parameter $\beta$, the
behavior of the electric field $\textbf{E}(\textbf{x})$ in Eq. (22)
is as follows:

\begin{equation}
\textbf{E}(\textbf{x})=
\begin{cases}
\dfrac{\tau\rho}{2\epsilon_{0}}\,\hat{\textbf{e}}_{\rho}-\dfrac{\tau^{3}\rho^{3}}{16\epsilon_{0}^{3}\beta^{2}}\,\hat{\textbf{e}}_{\rho}+{\cal
O}(\beta^{-4});\;\;\;\;&\rho < R,\vspace{0.3cm}\\ \dfrac{\tau
R^{2}
}{2\epsilon_{0}\rho}\,\hat{\textbf{e}}_{\rho}-\dfrac{\tau^{3}R^{6}}{16\epsilon_{0}^{3}\beta^{2}\rho^{3}}\,\hat{\textbf{e}}_{\rho}+{\cal
O}(\beta^{-4});\;\;\;&\rho
> R.
\end{cases}
\end{equation}
Hence, for the large values of $\beta$, the electric field
$\textbf{E}(\textbf{x})$ in Eq. (23) becomes the electric field of
an infinitely long cylinder in Maxwell electrostatics. If we
substitute Eq. (22) into Eq. (15), we will obtain the electrostatic
energy density for an infinitely long cylinder in Born-Infeld
electrostatics as follows:

\begin{equation}
u(\textbf{x})=
\begin{cases}
\epsilon_{0}\beta^{2}\Big(\sqrt{1+\big(\dfrac{\tau\rho}{2\epsilon_{0}\beta}\big)^{2}}-1\Big);\;\;\;\;\;\;&\rho < R,\vspace{0.2cm}\\
\epsilon_{0}\beta^{2}\Big(\sqrt{1+\big(\dfrac{\tau
R^{2}}{2\epsilon_{0}\beta\rho}\big)^{2}}-1\Big) ;\;\;\;\;&\rho
> R.
\end{cases}
\end{equation}

Using Eq. (24), the stored electrostatic energy per unit length for
an infinitely long cylinder in the radial interval $0 \leq\rho\leq
\Lambda\;(\Lambda> R) $ is given by

\begin{eqnarray}
\dfrac{U}{L}&=&
2\pi\epsilon_{0}\beta^{2}\Bigg\lbrace\int^{R}_{0}\Big(\sqrt{1+\big(\dfrac{\tau\rho}{2\epsilon_{0}\beta}\big)^{2}}-1\Big)\rho
d\rho + \int^{\Lambda}_{R}\Big(\sqrt{1+\big(\dfrac{\tau
R^{2}}{2\epsilon_{0}\beta\rho}\big)^{2}}-1\Big)\rho
d\rho \Bigg\rbrace\nonumber \\
&=& 2\pi\epsilon_{0}\beta^{2}\Bigg\lbrace-\dfrac{\Lambda^{2}}{2}
+\big(\dfrac{2\epsilon_{0}\beta}{\tau\sqrt{3}}\big)^{2}\bigg[\Big(1+\big(\dfrac{\tau
R}{2\epsilon_{0}\beta}\big)^{2}\Big)^{\frac{3}{2}}-1\bigg]+\dfrac{\Lambda^{2}}{2}\sqrt{1+\big(\dfrac{\tau
R^{2}}{2\epsilon_{0}\beta \Lambda}\big)^{2}}  \nonumber \\  \qquad
&-& \dfrac{R^{2}}{2}\sqrt{1+\big(\dfrac{\tau R}{2\epsilon_{0}\beta
}\big)^{2}} +\frac{1}{2}\big(\dfrac{\tau R^{2}}{2\epsilon_{0}\beta
}\big)^{2}\;\ln
\Bigg(\dfrac{\;\Lambda+\Lambda\sqrt{1+\big(\dfrac{\tau
R^{2}}{2\epsilon_{0}\beta \Lambda}\big)^{2}}}{R+R
\sqrt{1+\big(\dfrac{\tau R}{2\epsilon_{0}\beta }\big)^{2}}}\Bigg)
\Bigg\rbrace.
\end{eqnarray}
In the limit of large $\beta$, the expression for $\dfrac{U}{L}$ in Eq. (25) can be expanded in powers of $\dfrac{1}{\beta^{2}}$ as follows:
\begin{equation}
\dfrac{U}{L}|_{_{large \; \beta}} = \dfrac{\pi \tau^{2}R^{4}}{4 \epsilon_{0}}(\dfrac{1}{4}+\ln \dfrac{\Lambda}{R})+{\cal
O}(\beta^{-2}).
\end{equation}
 The first term on the right-hand side of Eq. (26) shows the stored electrostatic energy per unit length for an infinitely long cylinder in the radial interval
$0 \leq\rho\leq
\Lambda\;(\Lambda> R) $ in Maxwell electrostatics.

\section{Summary and Conclusions}
In 1934, Born and Infeld introduced a nonlinear generalization of
Maxwell electrodynamics, in which the classical self-energy of a
point charge like electron became a finite value [1]. We showed
that, in the limit of large $\beta$, the modified Gauss's law in
Born-Infeld electrostatics is
\begin{equation}
\oint_{S}\bigg[1+\dfrac{1}{2}\dfrac{\textbf{E}^{2}(\textbf{x})}{\beta^{2}}+\dfrac{3}{8}\dfrac{\big({\textbf{E}^{2}(\textbf{x})}\big)^{2}}{\beta^{4}}+{\cal
O}(\beta^{-6})\bigg]\textbf{E}(\textbf{x}).\textbf{n}\;da=\dfrac{1}{\epsilon_{0}}\int_{V}
\rho(\textbf{x})d^{3}x.
\end{equation}
By using the modified Gauss's law in Eq. (14), we calculated the
electric field of an infinite charged line and an infinitely long
cylinder in Born-Infeld electrostatics. The stored electrostatic
energy per unit length for the above configurations of charge
density has been calculated in the framework of Born-Infeld
electrostatics. Born and Infeld attempted to determine $\beta$ by
equating the classical self-energy of the electron in their theory
with its rest mass energy. They obtained the following numerical
value for the nonlinearity parameter $\beta$ [1]:
\begin{equation}
\beta=1.2\times10^{20}\; \dfrac{V}{m}.
\end{equation}
In 1973, Soff, Rafelski and Greiner [22] have estimated a lower
bound on $\beta$. This lower bound on $\beta$ is
\begin{equation}
\beta\geq 1.7\times10^{22}\; \dfrac{V}{m}.
\end{equation}
Resent studies on photonic processes in Born-Infeld theory show
that the numerical value of $\beta$ is close to
$1.2\times10^{20}\; {V}/{m}$ in Eq. (28) [23]. In order to obtain
a better understanding of nonlinear effects in Born-Infeld
electrostatics, let us estimate the numerical value of the second
term on the right-hand side of Eq. (19). For this purpose, we
rewrite Eq. (19) as follows:
\begin{equation}
\textbf{E}(\textbf{x})=\textbf{E}_{0}(\textbf{x})+\Delta\textbf{E}(\textbf{x})+{\cal
O}(\rho^{-5}),
\end{equation}
where
\begin{eqnarray}
\textbf{E}_{0}(\textbf{x})&:=&\dfrac{\lambda}{2\pi\epsilon_{0}\rho}\;\hat{\textbf{e}}_{\rho},\\
\Delta\textbf{E}(\textbf{x})&:=&-\dfrac{\lambda^{3}}{16\pi^{3}\epsilon_{0}^{3}\beta^{2}\rho^{3}}\;\hat{\textbf{e}}_{\rho}.
\end{eqnarray}

Using Eqs. (31) and (32), the ratio of
$\Delta\textbf{E}(\textbf{x})$ to $\textbf{E}_{0}(\textbf{x})$ is
given by
\begin{equation}
\dfrac{|\Delta\textbf{E}(\textbf{x})|}{|\textbf{E}_{0}(\textbf{x})|}=\dfrac{1}{2}\;\dfrac{\textbf{E}_{0}^{2}(\textbf{x})}{\beta^{2}}.
\end{equation}
Let us assume the following approximate but realistic values [24]:
\begin{equation}
L=1.80\; m, \;\;\;\rho=0.10\; m,\;\;\;Q=+24 \;\mu C,\;\;\;
\lambda=1.33\times 10^{-5}\;\frac{C}{m}\;.
\end{equation}
By putting Eqs. (28) and (34) into Eq. (33), we get
\begin{equation}
|\Delta\textbf{E}(\textbf{x})|\approx 2 \times
10^{-28}\;|\textbf{E}_{0}(\textbf{x})|.
\end{equation}
Finally, if we put Eqs. (29) and (34) in Eq. (33), we obtain
\begin{equation}
|\Delta \textbf{E}(\textbf{x})|\;\lesssim
10^{-32}\;|\textbf{E}_{0}(\textbf{x})|.
\end{equation}

In fact, as is clear from Eqs. (35) and (36), the nonlinear
corrections to electric field in Eq. (19) are very small for weak
electric fields. The authors of Ref. [25] have suggested a
nonlinear generalization of Maxwell electrodynamics. In their
generalization, the electric field of a point charge is singular
at the position of the point charge but the classical self-energy
of the point charge has a finite value. Recently, Kruglov [26,27]
has proposed two different models for nonlinear electrodynamics.
In these models, both the electric field of a point charge at the
position of the point charge and the classical self-energy of the
point charge have finite values. In future works, we hope to study
the problems discussed in this research from the viewpoint of
Refs. [25-27].


\section*{Acknowledgments}
We would like to thank the referees for their careful reading and
constructive comments.

\section*{Appendix A: A Generalized Action Functional for Abelian Born-Infeld Model with an Auxiliary Scalar Field}
\renewcommand{\theequation}{A.\arabic{equation}}
\setcounter{section}{0} \setcounter{equation}{0}

Let us consider the following action functional:
\begin{equation}
S(A,\psi)=\dfrac{1}{c}\int^{t}_{t_{0}}\int_{\mathbb
R^{3}}\bigg[\epsilon_{0}\beta^{2}\bigg(1-\omega_{1}\;\psi^{\lambda_{1}}(1+\dfrac{c^{2}}{2\beta^{2}}F_{\mu\nu}F^{\mu\nu})-\omega_{2}\;\psi^{\lambda_{2}}\bigg)-J^{\mu}A_{\mu}\bigg]d^{4}x,
\end{equation}
where $\psi$ is an auxiliary scalar field and
$\omega_{1},\;\lambda_{1},\;\omega_{2}$ and $\lambda_{2}$ are four
non-zero constants. The variation of Eq. (A.1) with respect to
$\psi$ and $A_{\mu}$ leads to the following classical field
equations:
\begin{equation}
\psi=\bigg[-\dfrac{\omega_{2}\;\lambda_{2}}{\omega_{1}\;\lambda_{1}}\bigg(1+\dfrac{c^{2}}{2\beta^{2}}F_{\mu\nu}F^{\mu\nu}\bigg)^{-1}\bigg]^{\dfrac{1}{\lambda_{1}-\lambda_{2}}},
\end{equation}
\begin{equation}
\partial_{\mu}(\psi^{\lambda_{1}}F^{\mu\nu})=\dfrac{\mu_{0}}{2\omega_{1}}J^{\nu}.
\end{equation}
Substituting Eq. (A.2) into  Eq. (A.3), we obtain the following
classical field equation:
\begin{equation}
\partial_{\mu}\bigg\lbrace\bigg[-\dfrac{\omega_{2}\;\lambda_{2}}{\omega_{1}\;\lambda_{1}}\bigg(1+\dfrac{c^{2}}{2\beta^{2}}F_{\alpha\gamma}F^{\alpha\gamma}
\bigg)^{-1}\bigg]^{\dfrac{\lambda_{1}}{\lambda_{1}-\lambda_{2}}}F^{\mu\nu}\bigg\rbrace=\dfrac{\mu_{0}}{2\omega_{1}}J^{\nu}.
\end{equation}
By choosing $\lambda_{1}=\lambda,\;\lambda_{2}=-\lambda $ and
$\omega_{1}=\omega, \;\omega_{2}=\dfrac{1}{4\omega} \;(\omega>0)$,
Eqs.(A.1) and (A.4) can be written as follows:
\begin{equation}
S(A,\psi)=\dfrac{1}{c}\int^{t}_{t_{0}}\int_{\mathbb
R^{3}}\bigg[\epsilon_{0}\beta^{2}\bigg(1-\omega\;\psi^{\lambda}(1+\dfrac{c^{2}}{2\beta^{2}}F_{\mu\nu}F^{\mu\nu})-\dfrac{1}{4\omega}\;\psi^{-\lambda}\bigg)-J^{\mu}A_{\mu}\bigg]d^{4}x,
\end{equation}
\begin{equation}
\partial_{\mu}\bigg(\frac{F^{\mu\nu}}{\sqrt{1+\frac{c^{2}}{2\beta^{2}}F_{\alpha\gamma}F^{\alpha\gamma}}}\bigg)=\mu_{0}J^{\nu}.
\end{equation}
Equation (A.5) is the generalized action functional for Abelian
Born-Infeld model with an auxiliary scalar field $\psi$. Also, Eq.
(A.6) is the inhomogeneous Born-Infeld equation (see Eq. (4)). If
we choose $\omega=\dfrac{1}{2}$ and $\lambda=1$ in  Eq. (A.5), we
will obtain the following action functional:
\begin{equation}
S(A,\psi)=\dfrac{1}{c}\int^{t}_{t_{0}}\int_{\mathbb
R^{3}}\bigg[\epsilon_{0}\beta^{2}\bigg(1-\dfrac{\psi}{2}(1+\dfrac{c^{2}}{2\beta^{2}}F_{\mu\nu}F^{\mu\nu})-\dfrac{1}{2\psi}\bigg)-J^{\mu}A_{\mu}\bigg]d^{4}x.
\end{equation}
The above action functional was presented by Tseytlin in his
studies on low energy dynamics of $D$-branes [28].

\section*{Appendix B: The Symmetric Energy-Momentum Tensor for Abelian Born-Infeld Model with an Auxiliary Scalar Field}
\renewcommand{\theequation}{B.\arabic{equation}}
\setcounter{section}{0} \setcounter{equation}{0}

In this appendix, we want to obtain the symmetric energy-momentum
tensor for Abelian Born-Infeld model with an auxiliary scalar
field. According to Eq. (A.5), the Lagrangian density for Abelian
Born-Infeld model with an auxiliary scalar field $\psi$ in the
absence of external source $J^{\mu}$ is
\begin{equation}
{\cal L} =
\epsilon_{0}\beta^{2}\Big(1-\omega\psi^{\lambda}\big(1+\dfrac{c^{2}}{2\beta^{2}}F_{\mu\nu}F^{\mu\nu}\big)-\dfrac{1}{4\omega}\psi^{-\lambda}
\Big).
\end{equation}
From Eq. (B.1), we obtain the following classical field equations:
\begin{equation}
\psi^{\lambda}=\frac{1}{2\omega}\Big(1+\dfrac{c^{2}}{2\beta^{2}}F_{\mu\nu}F^{\mu\nu}
\Big)^{-\frac{1}{2}},
\end{equation}
\begin{equation}
\partial_{\mu}\Big(\psi^{\lambda}F^{\mu\nu}\Big)=0.
\end{equation}
The canonical energy-momentum tensor for Eq. (B.1) is
\begin{eqnarray}
T^{^{\alpha}}_{\;\gamma}&=& \dfrac{\partial{\cal
L}}{\partial(\partial_{\alpha}A_{\eta})}(\partial_{\gamma}A_{\eta})+\dfrac{\partial{\cal
L}}{\partial(\partial_{\alpha}\psi)}(\partial_{\gamma}\psi)-\delta^{\alpha}_{\,\gamma}\cal
L\nonumber \\
&=& -2\epsilon_{0}c^{2}\omega
\psi^{\lambda}F^{\alpha\eta}(\partial_{\gamma}A_{\eta})-
\epsilon_{0}\beta^{2}\delta^{\alpha}_{\,\gamma}\Big(1-\omega
\psi^{\lambda}\big(1+\dfrac{c^{2}}{2\beta^{2}}F_{\mu\nu}F^{\mu\nu}\big)-\dfrac{1}{4\omega}\psi^{-\lambda}\Big).
\end{eqnarray}
 Using Eqs. (B.2) and (B.3), the canonical energy-momentum tensor $T^{^{\alpha}}_{\;\gamma}$
in Eq. (B.4) can be rewritten as follows:
\begin{eqnarray}
T^{^{\alpha}}_{\;\gamma}&=&-\epsilon_{0}c^{2}\Big(1+\dfrac{c^{2}}{2\beta^{2}}F_{\mu\nu}F^{\mu\nu}
\Big)^{-\frac{1}{2}}F^{\alpha\eta}F_{\gamma\eta}\nonumber
\\ \qquad
&\;&
-\epsilon_{0}\beta^{2}\delta^{\alpha}_{\;\gamma}\Big(1-\big(1+\dfrac{c^{2}}{2\beta^{2}}F_{\mu\nu}F^{\mu\nu}\big)^{\frac{1}{2}}\Big)+\partial_{\eta}M^{\eta\alpha}_{\;\;\;\gamma},
\end{eqnarray}
where
\begin{equation}
M^{\eta\alpha}_{\;\;\;\gamma}:=
\dfrac{1}{\mu_{0}}\dfrac{F^{\eta\alpha}}{\sqrt{1+\frac{c^{2}}{2\beta^{2}}F_{\mu\nu}F^{\mu\nu}}}A_{\gamma},\quad
M^{\alpha\eta}_{\;\;\;\gamma}=-M^{\eta\alpha}_{\;\;\;\gamma} .
\end{equation}
After dropping the total divergence term
$\partial_{\eta}M^{\eta\alpha}_{\;\;\;\gamma}$ in Eq. (B.5), we get
the following expression for the symmetric energy-momentum tensor:
\begin{equation}
T^{^{\alpha}}_{\;\gamma}=-\epsilon_{0}c^{2}\Big(1+\dfrac{c^{2}}{2\beta^{2}}F_{\mu\nu}F^{\mu\nu}
\Big)^{-\frac{1}{2}}F^{\alpha\eta}F_{\gamma\eta}-\epsilon_{0}\beta^{2}\delta^{\alpha}_{\;\gamma}\Big(1-\big(1+\dfrac{c^{2}}{2\beta^{2}}F_{\mu\nu}F^{\mu\nu}\big)^{\frac{1}{2}}\Big).
\end{equation}
If we use Eqs. (6) and (B.7), we will obtain the electrostatic
energy density for Abelian Born-Infeld model with an auxiliary
scalar field as follows:
\begin{equation}
u(\textbf{x})=
T^{0}_{\;0}(\textbf{x})=\epsilon_{0}\beta^{2}\Big(\frac{1}{\sqrt{1-\frac{\textbf{E}^{2}(\textbf{x})}{\beta^{2}}}}-1\Big).
\end{equation}

\end{document}